# Representative Methods of Computational Socioeconomics

Tao Zhou

CompleX Lab, University of Electronic Science and Technology of China, Chengdu 611731, People's Republic of China
**E-mail**: zhutou@ustc.edu

**Abstract**

The increasing data availability and imported analyzing tools from computer science and physical science have sharply changed traditional methodologies of social sciences, leading to a new branch named computational socioeconomics that studies various phenomena in socioeconomic development by using quantitative methods based on large-scale real-world data. Sited on recent publications, this Perspective will introduce three representative methods: (i) natural data analyses, (ii) large-scale online experiments, and (iii) integration of big data and surveys. This Perspective ends up with in-depth discussion on the limitations and challenges of the above-mentioned emerging methods.

## 1. Introduction

    Social sciences study the social structure based on activities of and relations between human beings, encompass a wide array of academic disciplines including sociology, economics, politics, linguistics, jurisprudence, and many other branches. In despite of the intrinsic difficulty of quantification, social sciences recently show higher and higher level of quantification and becomes increasingly dependent on real data [1-5]. However, the traditional way to obtain real data has many limitations. For example, survey data from questionnaires usually contains insufficient samples for solid statistics and suffers from social desirability bias [6], while large-scale and precise censuses consume huge resources and lack timeliness.

    Fortunately, the ongoing digital wave results in unprecedentedly rich data that records microscopic processes of social and economic development, including satellite remote sensing data, mobile phone data, social media data, and so on. These data are usually of large sizes, almost in real time and with high resolution, which can reduce the small-sample bias and the invisible parts in the developing processes. Leveraging these new data sources, we can in principle make great progress in perceiving socioeconomic status, evaluating and amending known theories, enlightening and creating new theories, detecting abnormal events, predicting future trends, and so on. Such opportunity has led to the emergence of a new branch named *computational socioeconomics* that studies various phenomena in socioeconomic development through analyzing large-scale real-world data by emplying quantitative methods, with a particular attention to economic development problems related to social processes and social problems related to economic development [7]. The computational socioeconomics can be considered as a new branch of socioeconomics resulted from the transformation of methodology, or as a new branch of computational social science by emphasizing on socioeconomic problems [2,8]. The main research problems in computational socioeconomics can be roughly laid out in three layers: (i) Global socioeconomic development, with a particular attention to inequality [9] and fitness [10] in economics development, as well as the global evolution of culture [11]; (ii) Regional economic development, with a particular attention to characterizing the regional economic state [12], revealing the hidden complexity of regional economic structure [13], and exploring effective ways towards better economics [14]; (iii) Individual socioeconomic status, mainly concentrating on how to quantify individual employment status, pecuniary condition, living standard and mental condition in an efficient and unobtrusive way [15,16].

    Built on a growing number of papers mainly published in the last decade, this Perspective classifies the novel data-driven methods into three representative categories: (i) natural data analyses, (ii) large-scale online experiments, and (iii) integration of big data and surveys. Each of the following three sections will introduce one method, providing the fundamental idea, one

typical research example and a few related works. This Perspective ends up with in-depth discussion on the limitations and challenges of the above-mentioned emerging methods.

## 2. Natural Data

Natural data refers to the data collected in a way that the targets are not aware of the fact that they are recorded and analyzed. Mobile phone data and social media data are typical examples. Natural data can best reflect the real behaviors and thus considered to be even more honest than the so-called *honest signals* [17] collected by long-worn sensors, because volunteers' behaviors will change to be more social desirable (sometimes inconsciently) if they know they are involved in some experiments.

A typical research example is how to quantify the religious segregation via analyzing social media data [18]. Religion locates in the core of culture, exhibiting significant positive values like to facilitate human cooperation [19] and to improve life satification [20]. At the same time, people with different faiths tend to form relatively isolated communities, which may result in misunderstanding, prejudice, animosity, extremism or even violence [21]. Therefore, to quantify the extent of religious segregation and to find out effective methods that can promote cross-religion communications are valuable. Questionnaire is not suitable in this scenario because some questions could cause offence. Instead, Hu *et al.* [18] collected publicly available data from weibo.com (one of the largest online social platforms in China, similar to Twitter), where users with religious beliefs (i.e., believers) are identified by NLP techniques (see details in [18]). The subgraph of the weibo follower-followee network, induced by these believers, is analyzed. Four major conclusions are drawn from such natural data. Firstly, users with different beliefs are largely segregated, and the religion network in weibo.com shows even higher level of segregation than well-known social networks with remarkable segregation like sexual partnerships mixed by races [22] and the Twitter follower-followee network of politicians in democratic party and republican party [23]. Secondly, cross-religion links play the most critical role in maintaining network connectivity, whose removal will lead to much faster breakdown of the network than links selected by other well-known centralities [24], such as betweenness [25] and bridgeness [26]. Thirdly, according to the comparison with null model [27], Buddhism contributes the most to cross-religion communications. Lastly, about half cross-religion links point to charitable nodes, suggesting that charity may be a common interest that can stride across the ideological barriers between religions.

The above work for the first time quantifies the religious segregation based on natural data. Recently, a growing number of works on socioeconomic issues have utilized natural data. For example, tweets of Chrisrtians and atheists are analyzed to show the value of religion in bringing happiness [28], searching queries in Google [29] and activity times in Twitter [30] are used to estimate unemployment, records in office automation systems are utilized to predict future promotion and resignation of employees [31], and online resumes of anonymous job seekers are used to quantify the differences in expected salaries of groups with different genders and statures, which can be considered as informative clues about potential discrimination in employment on gender and stature [32].

## 3. Online Experiments

Although natural data is powerful in revealing underlying reality, we should be aware of that behavioral experiments are still very important in social sciences, because well-designed experiments could eliminates the impacts from confounding factors and unobserved variables and thus gain causal relationships, which are usually more meaningful and relevant than uninterpreted correlations obtained from data mining techniques [33]. However, experimental studies are largely limited by the high costs in volunteer recruitment and experimental implementation. Therefore, the samples are usually too few to draw statistically solid conclusions. Sometimes, to save costs and/or make the implementation more convinient, researchers directly recruit volunteers inside campuses or even from their classes, resulting in unrepresentative samples. In fact, results from behavioral experiments are often nonreplicable. Recently, the Open Science Collaboration repeated 100 psychological experiment and reported a very low rate ≤40% of successful reproduction [34], Camerer *et al.* [35,36] replicated 18 laboratory experiments in economics published in *American Economic Review* and *Quarterly Journal of Economics* in 2011-2014, as well as 21 experimental studies in social sciences published in *Nature* and *Science* in 2010-2015. In the former case, they reported a significant effect in the same direction as the original study for 11 replications (61%) with on average the replicated effect size being 66% of the original [35], while in the latter case these two numbers are 13 (62%) and 50% [36].

Using online platforms to design and implement large-scale behavioral experiments may overcome the above-mentioned shortcomes, such as high costs, insufficient samples and low representativeness. Bond *et al.* [37] designed an online political mobilization experiment during the 2010 US congressional elections, which attracted 61 million Facebook users. In the experiment, all users of at least 18 years old in the US who accessed the Facebook website on 2 November 2010 (i.e., the day of the US congressional elections) were randomly assigned to a social message group, an informational group or a control group. Each user in the social message group was shown a statement at the top of the *News Feed*, which encouraged the user to vote, provided a link to find local polling places, showed a clickable button reading *I Voted*, showed a counter indicating how many other Facebook users had previously reported voting, and displayed up to six randomly selected profile pictures of the user's Facebook friends who had already clicked the *I Voted* button. The informational message group was shown the message but they were not shown any faces of friends. The control group did not receive any message at the top of their *News Feed*. The experimental results showed that users in the social message group were 2.08% more likely to click on the *I Voted* button than those in the informational message group, that is, the socialized environment could promote political self-expression. In addition, users receiving relevant information in their *News Feed* are 0.39% more likely to vote than those in the control group. Recalling the 2000 US presidential election where Bush beat Gore in Florida by only 537 votes (<0.01% of total votes in Florida), 0.39% is already a significant difference in election. In a word, with the help of Facebook, Bond *et al.* successfully implemented an online experiment with inconceivably large number of participants, and showed a clear evidence that online political mobilization can essentially contribute to real-world behaviors.

In addition to co-designing experiments and co-developing necessary software (like [37]), researchers could collaborate with online platforms in various ways. Salganik *et al.* [38] used Internet to recruit volunteers and to launch the experimental software. They attracted 14341 volunteers to participate in the music evaluation experiment, which suggested the existence of a strong Matthew effect (rich gets richer). Pichard *et al.* [39] proposed a mechanism to allocate rewards to all contributors for a task, and then utilized online social networks to propagate their idea and generate the incentive network. They quickly built a team with about 4400 members and eventually won the *DARPA Network Challenge 2009* that asks for the coordinates of 10 red weather balloons placed at different locations in the continental United States. Van de Rijt *et al.* [40] directly used existing functions of online platforms (e.g., crowdfunding website Kickstarter, online product evaluation website Epinions, etc.) to implement their experiments (e.g., to randomly choose 100 projects and donate a small percentage of funding goal to them, to evaluate randomly sampled products as *Very Helpful*, etc.), revealing a sharp success-breeds-success phenomenon. Pennycook *et al.* [41] sent private messages to 5379 selected users from Twitter and recruited thousands of volunteers in Amazon Mechanical Turk to investigate why people share misinformation, and the experimental results suggested that sharing does not necessarily indicate belief but subtly shifting attention to accuracy will increase the quality of news that people subsequently share.

## 4. Survey Expanding

The increasing data availability provides us unprecedented opportunity to collect and analyze population-scale data, such as social media data from Facebook, Twitter, Wechat and Weibo, and call detail records and mobility trajectories from smart phones. Although the cost to obtain such data is much lower than that of population-scale censuses or surveys, we cannot use these data to directly answer important questions concerning household incomes, mental and physical health, social segregation and social mobility, employment status, and so on. Therefore, we call these data as *easy-to-access indirect data*. Let's consider a certain study that requires population-scale data on household incomes. Obviously, it is very hard to know the household income of every family because a poor country cannot support a population-scale economic census, and even if the census is completed, the microscopic data is usually confidential and not open for public or research institutions. Under the circumstances, we can obtain household incomes of a tiny number of families via routine surveys. This much smaller data can be used as training data, based on which we can apply machine learning techniques to predict household income of a family based on easy-to-access indirect data, such as the mobile phone data of the family members [42]. Such a small-size data set directly pointing to the research target is called *hard-to-get direct data*. Combining the population-scale easy-to-access indirect data, a small sample of hard-to-get direct data, and a properly selected or well-designed algorithm to infer the missing direct data of individuals other than the sample is a novel and powerful methodology. Although the inferred data is not perfect, it can be obtained at a very low cost and its accuracy is usually sufficient for studies.

Using the above methodology, Blumenstock *et al.* [9] produced the wealth map of Rwanda at a very high resolution, so that poor areas and individuals can be easily identified. Collaborated with telecommunications operators, Blumenstock *et al.* obtained 1.5 million mobile phone users' call detail records including billions of calls and short messages. They further recruited 856 volunteers from those mobile phone users and asked volunteers to fill in detailed questionnaires about their socioeconomic statuses. They trained a machine learning algorithm to predict the 856 volunteers' wealth and other socioeconomic indicators, and then generated the out-of-sample prediction for the 1.5 million users. They found a strong correlation between the government ground truth data and the predicted wealth data after aggregating them to the district level. At a very low cost (only about 0.06% of mobile users are sampled to fill in questionnaires), their method could map the distribution of wealth and other socioeconomic indicators for the full national population.

In addition to socioeconomic status, this methodology can also be applied in analyzing mental health conditions at the individual level. Although mental problems of specific individuals do not belong to the scope of social sciences, the temporal trends of anxiety level, depression level and suicide rate of a society, and the relationships between the overall prevalence of mental diseases and political system, economic development and educational standards are highly concerned issues in social sciences. Taking depression as example, De Choudhury et al. collected 476 depression patients' health records and their one-year Twitter data before onsets [43], as well as 165 baby mothers' health records and their Facebook data [44]. De Choudhury et al. [43,44] respectively built machine learning models to predict depression and postpartum depression, and the accuracy of the former can approach to 70%. In a word, they used a few hundreds of health records (hard-to-get direct data) and corresponding Twitter and Facebook data (easy-to-access indirect data) to train their machine learning models, and in principle their models can be applied to the entire Twitter data to estimate the overall level of depression of a country or region, or a certain group of people with predefined demographics. Analogously, Weibo text [45] and Instagram photos [46] can also be utilized to identify depression patients at the early stage.

## 5. Discussion

Thanks to the big data and artificial intelligence techniques, papers applying the three highlighted methods appear more and more frequently in the last decade, which are leading a profound and irreversible reform of routine methodologies of social sciences [5,7]. At the same time, there are still some apparent limitations of these promising methodologies, which are worth special attentions in the future studies.

In brief, a theory in natural science has three progressive abilities: (i) it could *explain* some observed phenomena, (ii) it could *predict* what will happen with given conditions, and (iii) it can be applied to *control* real systems. This explain-predict-control chain constitutes a common validation process for a newly-proposed theory. In social sciences, we cannot find or construct an idea system with every component and ingredient being explicitly and quantitatively known, and the functions and effects of a theory cannot be completely distinguished from others. Therefore, we cannot expect to understand the role of a social theory like the role of the gravity law in artificial satellites or the role of the general relativity in global position systems. Accordingly, we do not have the illusion that a social theory can be well utilized to control a real system, yet we expect at least it can be applied to intervene a system. However, in order to go from predicting to intervening, we have to know the relevant causal relationships, not just the correlations [33]. For example, although we know the temporal orderliness of campus lifestyle (e.g., the time to take shower) is significantly correlated with academic performance [47], and this correlation (together with some other features) can be utilized to predict academic performance with an acceptable accuracy [48], we cannot improve a student's academic performance by simply rescheduling her/his taking shower time. It is because a regular schedule of showers is not a causation of a high GPA score, while there may be some unobservable causations (e.g., conscientiousness [49]) that lead to both high/low orderliness and high/low GPA scores. Online experiments are able to reveal clear causality and sometimes experiments themselves do intervene the real world as Bond et al. did [37]. In comparison, statistical techniques that evaluate the intervention effect (e.g., propensity score matching [50] and difference in differences [51]) and discover the causal knowledge (e.g., asymmetric Shapley values [52] and causal Shapley values [53]) could provide valuable clues, but those inferred results are less confident than experimental results. So generally speaking, we are still far from the ability of intervention. Notice that, more and more data sets with time stamps are available, which are not exactly the same to but similar to longitudinal data, therefore researchers should explore how to apply, extend and develop known tools and methods for longitudinal data [54] to deal with socioeconomic problems.

The last section introduces how big data and traditional questionnaires can be combined to get added value. In the combination, questionnaires are not changed. That is to say, questionnaires are not improved by big data. What we expect in the next step is that big data can help in designing better questionnaires. On one hand, routine method to design a questionnaire may miss important items, which are probably found out in advance by analyzing correlations between candidate factors and targets with the help of big data and data mining techniques. On the other hand, the gap between correlation and causality can be bridged by a properly-designed questionnaire. In this way, the novel data-driven methodologies and traditional methodologies are deep integrated. A very few works have already seriously investigated the feasibility and possible pathways [55]. In addition, variable importance methods that are currently used to make machine learning interpretable can also be used to understand what variables are important to a certain outcome, and this understanding may lead to a better design of questionnaires. For example, Shapley values, coming from coalition game theory [56], has been shown to be a very good measure of global variable importance [57,58], and thus may be applied in designing questionnaires using knowledge gained from big data.

Lastly, the methodologies introduced in this Perspective may give rise to a number of ethical issues, which must be cautiously treated. Firstly, the private information needs to be strictly protected. When natural data is used in a study, the involved individuals are not aware of the fact that they are under investigation. If the data comes from unpublicized systems, it is clear that the privacy should be protected. Even if the data is collected from public websites, the private information has to be protected because individuals probably do not want other people knowing the analytic results related to them, or do not want to see the results themselves. For example, a user is willing to share his information to Facebook friends, which does not imply that he agrees to be known as a predicted depression patient [43] or a predicted gay [59] based on his shared data. Furthermore, even if a data set is anonymous, it is possibly to be de-anonymized [60-62]. Therefore, we should be very carefully in reporting analytic results and sharing scientific data to avoid disclosure of private information. Secondly, in some online experiments, to capture the real responses of subjects, they are kept to be unaware of the experiments. Therefore, researchers have to carefully evaluate in advance whether the materials shown to subjects (e.g., rumours, violence information, extreme and fanatical views, etc.) and the feedback from researchers (in particular the negative feedback against some selected subjects) will result in long-standing negative impacts on subjects' emotion and mentality. Thirdly, both large-scale data and survey data often contain representation bias, that is, a data set is representative of its source but may not be representative of other ensembles [63]. For example, users randomly sampled from a social media website probably over-represents young people while under-represents aged people in a population. Therefore, we should be cautious when applying the trained models or extending the results to the population different from the original source. An impressive lesson is that a well-performed classifier trained on ImageNet is not good at classifying images from India because only about 2% of images in ImageNet are from India [64]. Although the individual awareness of ethical issues is important and helpful, institutions and funders (may through the Institutional Review Boards) ought to play a vital role in developing informed regulatory framework and ethical guidance for researchers, and evaluating potential privacy and ethical concerns before studies were started [8].

## Acknowledgements

This work is partially supported by the National Natural Science Foundation of China under Grant No. 11975071.